\documentclass[conference]{IEEEtran}
\usepackage{graphicx}
\usepackage{multirow}
\usepackage{amssymb}
\usepackage{array}
\usepackage{booktabs}
\usepackage{tabularx}
\usepackage{wasysym}

\usepackage{hyperref}
\usepackage[T1]{fontenc}

\usepackage[whole]{bxcjkjatype}  

\usepackage{algorithm}
\usepackage{algcompatible}
\usepackage[noend]{algpseudocode}
\usepackage[normalem]{ulem}
\usepackage{hyperref}

\usepackage{footnote}
\makesavenoteenv{tabular}
\makesavenoteenv{table}

\usepackage{threeparttable}

\algnewcommand{\LeftComment}[1]{\Statex \(\triangleright\) #1}

\usepackage{color}

\newenvironment{mybullet}{\begin{list}{$\bullet$}
    {\setlength{\topsep}{0mm}\setlength{\itemsep}{0mm}
      \setlength{\parsep}{0.5mm}
      \setlength{\itemindent}{0mm}\setlength{\partopsep}{0mm}
      \setlength{\labelwidth}{15mm}
      \setlength{\leftmargin}{8mm}}}{\end{list}}

\usepackage{framed}

\begin{document}

\title{
\begin{framed}
    \normalsize
    \raggedright
    If you cite this paper, please use the following reference:\\Ryo Kawaoka, Daiki Chiba, Takuya Watanabe, Mitsuaki Akiyama, and Tatsuya Mori ``A First Look at COVID-19 Domain Names: Origin and Implications,'' {\it In Proceedings of the Passive and Active Measurement Conference 2021} (PAM 2021)
\end{framed}
\leavevmode
A First Look at COVID-19 Domain Names: \\
Origin and Implications}


\makeatletter
\newcommand{\linebreakand}{%
  \end{@IEEEauthorhalign}
  \hfill\mbox{}\par
  \mbox{}\hfill\begin{@IEEEauthorhalign}
}
\makeatother

\author{
\IEEEauthorblockN{Ryo Kawaoka}
\IEEEauthorblockA{\textit{Waseda University} \\
Tokyo, Japan \\
k-hsw119@nsl.cs.waseda.ac.jp}
\and
\IEEEauthorblockN{Daiki Chiba}
\IEEEauthorblockA{\textit{NTT Secure Platform Laboratories} \\
Tokyo, Japan \\
daiki.chiba@ieee.org}
\and
\IEEEauthorblockN{Takuya Watanabe}
\IEEEauthorblockA{\textit{NTT Secure Platform Laboratories} \\
Tokyo, Japan \\
watanabe@nsl.cs.waseda.ac.jp}
\linebreakand 
\IEEEauthorblockN{Mitsuaki Akiyama}
\IEEEauthorblockA{\textit{NTT Secure Platform Laboratories} \\
Tokyo, Japan \\
akiyama@ieee.org}
\and
\IEEEauthorblockN{Tatsuya Mori}
\IEEEauthorblockA{\textit{Waseda University/NICT/RIKEN AIP} \\
Tokyo, Japan \\
mori@seclab.jp}
}

\maketitle

\begin{abstract}

This work takes a first look at domain names related to COVID-19 (Cov19doms in short), using a large-scale registered Internet domain name database, which accounts for 260M of distinct domain names registered for 1.6K of distinct top-level domains. We extracted 167K of Cov19doms that have been registered between the end of December 2019 and the end of September 2020. 
We attempt to answer the following research questions through our measurement study: {\bf RQ1:} {\em Is the number of Cov19doms registrations correlated with the COVID-19 outbreaks?}, {\bf RQ2:} {\em For what purpose do people register Cov19doms?} 
Our chief findings are as follows: (1) Similar to the global COVID-19 pandemic observed around April 2020, the number of Cov19doms registrations also experienced the drastic growth, which, interestingly, pre-ceded the COVID-19 pandemic by about a month, 
(2) 70 \% of active Cov19doms websites with visible content provided useful information such as health, tools, or product sales related to COVID-19, and 
(3) non-negligible number of registered Cov19doms was used for malicious purposes. 
These findings imply that it has become more challenging to distinguish domain names registered for legitimate purposes from others and that it is crucial to pay close attention to how Cov19doms will be used/misused in the future.


\end{abstract}
\section{Introduction}
\label{sec:intro}

Several researchers have conducted Internet measurement studies to understand how the COVID-19 pandemic affected the Internet and user behaviors~\cite{Favale_2020,c2020impact,IMC20-covid-isp,IMC20-covid-mobile,IMC20-facebook}.
Favale et al. and Feldmann et al.~\cite{Favale_2020,IMC20-covid-isp} explored the changes in Internet traffic, Lutu et al.~\cite{IMC20-covid-mobile} explored the changes in traffic and its impact on user mobility in mobile operators, Candela et al.~\cite{c2020impact} analyzed the impact of Internet traffic changes on network latency, and Boettger et al.~\cite{IMC20-facebook} analyzed the changes in social media access patterns and the implications.
The details of these studies will be discussed in Section~\ref{sec:related}.

To the best of our knowledge, there has been no academic study that has analyzed the impact of COVID-19 in terms of registered domain names.
This work takes a first look at domain names related to COVID-19 (Cov19doms in short), using a large-scale set of registered domain names.
We note that the only literature we have been able to find on this subject is a blog article~\cite{cyberthreatcoalition2}, which analyzed the domain names associated with COVID-19.
The article reported that the number of COVID-19 domain name registrations has spiked in mid-March 2020, with some days seeing the registration of more than 5,000 Cov19doms.
However, we found that the data used in the article contained many false positives due to the naive string match heuristics. 
Also, this data is no longer updated since May 2020, so we cannot perform a longer-term analysis using the data.
In this study, we attempt to extract Cov19doms accurately and analyze how it changes over a long period of time.

With so many of us keeping an eye on COVID-19 and spending more and more of our time online, it is crucial to understand the origins and implications of Cov19doms.
Given these backgrounds in mind, we attempt to answer the following research questions:
\begin{mybullet}
\item[{\bf RQ1}:] {\em Is the number of Cov19doms registrations correlated with the COVID-19 outbreaks?} 
\item[{\bf RQ2}:] {\em For what purpose do people register Cov19doms?}
\end{mybullet}

To address the research questions, we compiled an exhaustive list of Cov19doms using a large-scale registered Internet domain name database~\cite{domainlistsio}, which accounted for 260M of distinct domain names registered for the 1.6K of top-level domains.
Using the dataset, we found that at least 167K of distinct Cov19doms containing strings such as ``covid'' or ``corona'' have been registered from the end of December 2019 to the end of September 2020. 
We attempt to study how domain name registration behavior changed with the emergence of COVID-19; i.e., we examine whether or not the time-series of COVID-19 infections is correlated with the time series of domain name registrations.

Next, from the 167K of Cov19doms, we extracted active websites that used Cov19doms by checking DNS A record and HTTP/HTTPS response.  
We then randomly sampled 10,000 of the Cov19doms websites to study how Cov19doms are used in the wild.
By applying cluster analysis to the screenshots, we systematically classified 10K websites.
For the remaining general websites, we performed manual inspection with the aid of three evaluators. 
We also leveraged online virus-testing services to check whether some Cov19doms were used for malicious activities.

Our chief findings are as follows: 
\begin{mybullet}
 \item Similar to the global COVID-19 pandemic observed around April 2020, the number of Cov19doms registrations also experienced drastic growth, which, surprisingly, preceded the COVID-19 pandemic by about a month. 
\item 70 \% of active Cov19doms websites with visible content provided useful information such as health, tools, or product sales related to COVID-19.
\item Non-negligible number (roughly 4\%) of registered Cov19doms have been used for malicious purposes such as phishing or malware distribution.
\end{mybullet}

These findings imply that it has become more challenging to distinguish between domain names registered for legitimate purposes and those that are not.
It was also indicated that it is necessary for researchers who analyze domain names, and even operators and blacklisters who take security measures based on domain names to pay close attention to how Cov19doms currently parked or in preparation will be used/misused in the future.

\section{Data}
\label{sec:data}

\subsection{Collecting Cov19doms}
To collect registered Cov19doms, we used a large-scale commercial domain name database, domainlist.io~\cite{domainlistsio}.
This database contains snapshots of approximately 260M domain names taken from 1.6K of different TLDs, and we continued to retrieve data daily from 27 December 2019 to 20 September 2020.
Of the $98,940,555$ domain names that have been newly registered since December 27, 2019,
we first extracted the domain names that contained ``covid'' or ``corona'' as a substring.
As a result, we obtained a total of $170,846$ Cov19doms.
We note that this approach could include false positives such as ``covideo.co.uk'', for instance.
However, we can safely ignore the effect of false positives in the following analysis, as our manual inspection of the randomly sampled data showed that the occurrence of such false positives was extremely rare as these words are.
We believe that these words, especially in the COVID-19 era, are mostly used in the context of a specific purpose, i.e., ``severe acute respiratory syndrome coronavirus 2,'' resulting in fewer false positives.

To study the characteristics of the Cov19doms, it is essential that we can get information about the creation date of the domain names. 
Therefore, we used the WHOIS information for the extracted Cov19doms to obtain information on the date and time the domain name was created.
If the creation date of a domain name was older than December 27, 2019, those domain names were excluded from the following analysis.
This resulted in a total of 166,825 Cov19doms, as shown in Table~\ref{tab:data_stats2}. To ensure that domains registered before December 27, 2019 were not related to COVID-19, we manually checked on them and found it be correct. In fact, most of them were related to  Coronado city in California, U.S.

We investigated where the specific words related to COVID-19, i.e., ``covid'' and ``corona'', are located in
the left-most labels of Cov19doms (e.g., ``covidcare'' in \texttt{covidcare[.]example}) and confirmed that (a) 59.6\% are at the beginning, (b) 24.2\% are at the end, and (c) 16.2\% are in the middle of them. 
The patterns (a) and (b) mean that the left-most labels of Cov19doms were generated by concatenating any character at the beginning or end of the COVID-19-related words such as ``covid''.
We believe that patterns (a) and (b) are less likely to cause false positives than pattern (c).
We further investigated the extent to which similar COVID-19-related words, ``covid'', ``covid19'', and ``covid-19'', are included in Cov19doms and found that they are 41,718, 32,671, and 10,120 Cov19doms, respectively. These numbers do not overlap, because we checked Cov19doms that contain ``covid19'' and ``covid-19'' earlier. 
It is interesting that ``covid19'' is more common in Cov19doms than its formal name of the desease, ``covid-19''. 
Among these, ``covid'' was most frequently included in Cov19doms, and as far as we manually checked, the majority of cases (about 40\%) were used in the context of the COVID-19.
One of the reasons why ``covid'' is included in Cov19doms in large numbers is that there are cases where various numbers are added to the end of ``covid'' (e.g., covid-2019, covid-2020, and covid-2021). We expect those domain names to have been acquired for speculative purposes.

We looked into what country registered Cov19doms firstest by usinig WHOIS registrant information. Of the 165,185 Cov19doms we extracted, 153,243 domains had valid WHOIS registrant country information. Among the countries, United States was the first to register Cov19doms. The top-5 countries registered Cov19doms were United States (85,970), Canada (17,229), Panama (6,781), Germany (4,533) and United Kingdom (4,237).

\subsection{Collecting Active Websites Using Cov19doms}
With the aim of studying the usage of Cov19doms, we extract the active websites that are operating using Cov19doms.
To extract active websites, we first check the DNS A record to determine if an IP address is assigned to the extracted Cov19doms. 
We then send an HTTP/HTTPS request to the domain name where the DNS A record exists, and record the response.
Specifically, we check if a connection can be established to Port 80 and Port 443 of each host that had a Cov19dom. Next, if a connection with either port can be established, we made an HTTP/HTTPS request to those hosts and checked whether the content could be retrieved from them.
This step removes websites that caused connection timeouts and/or TLS errors such as invalid certificate.
These steps resulted in a total of 77,333 of active websites that use Cov19doms, as shown in Table~\ref{tab:data_stats2}.

\begin{table}[tbp]
     \centering
     \normalsize
     \caption{Statistics of extracted Cov19doms data.}
     \label{tab:data_stats2}
     \begin{tabular}{l|r}
     \hline
                & \# of domain names \\
     \hline\hline
    Orig. Cov19doms & 170,846 \\
    WHOIS check & 166,825 \\
    DNS check & 144,522 \\
    HTTP/HTTPS check & 77,333 \\
     \hline
     \end{tabular}
\end{table}

\section{Measurement Study}
\label{sec:measurement}

Figure~\ref{fig:measurement_overview} presents an overview of the measurement processes.
We first study the correlation between the number of COVID-19 infections and the number of Cov19doms registrations (Sec~\ref{sec:correlation}).
For this analysis, we used the statistics on the number of COVID-19 infections by country, provided by WHO~\cite{who-covid19}.
We then study how Cov19doms are used for various websites (Sec~\ref{sec:categorization}).
The classification of active websites operated using Cov19doms was manually performed by three evaluators.
Due to the large number of websites to be analyzed, we conducted a random sampling study.
Finally, we report the analysis of Cov19doms that have been used for malicious activities (Sec~\ref{sec:malicious}). 
We used VirusTotal~\cite{virustotal} to investigate the presence of malicious sites using Cov19doms.

\begin{figure*}[!t]
    \centering
    \includegraphics[width=0.8\linewidth]{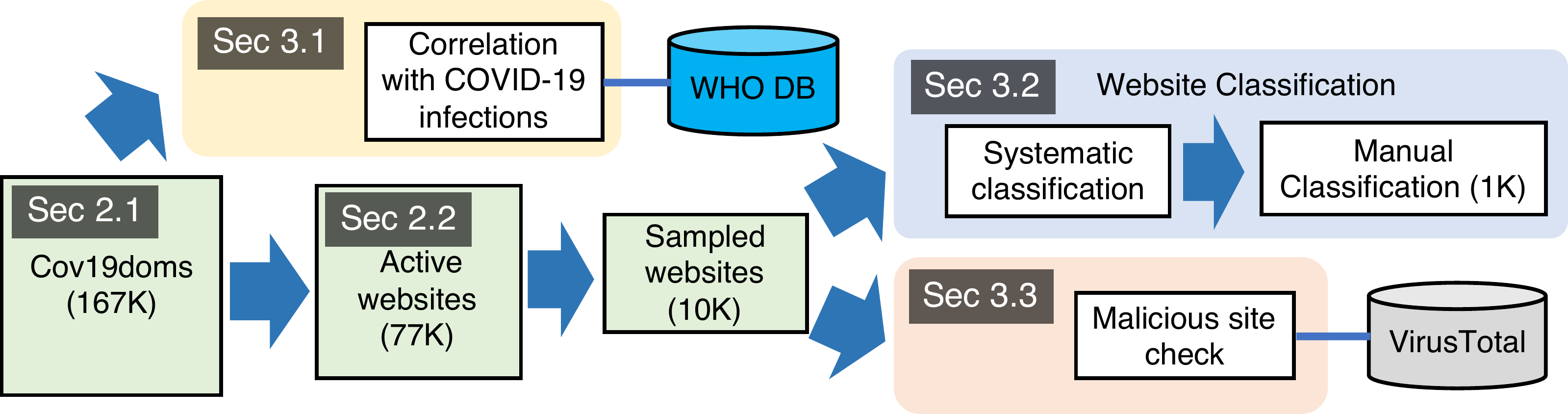}
    \caption{Overview of the measurement processes.}
    \label{fig:measurement_overview}
\end{figure*}

\subsection{Number of new infections and Cov19doms registrations}
\label{sec:correlation}
We analyze the online behavior of people around the world in response to the unprecedented event of COVID-19 through the lens of DNS.
Specifically, we examine whether or not the time series of COVID-19 infections is correlated with the time series of domain name registrations.

\begin{figure*}[tbp]
    \centering
    \includegraphics[width=0.49\linewidth]{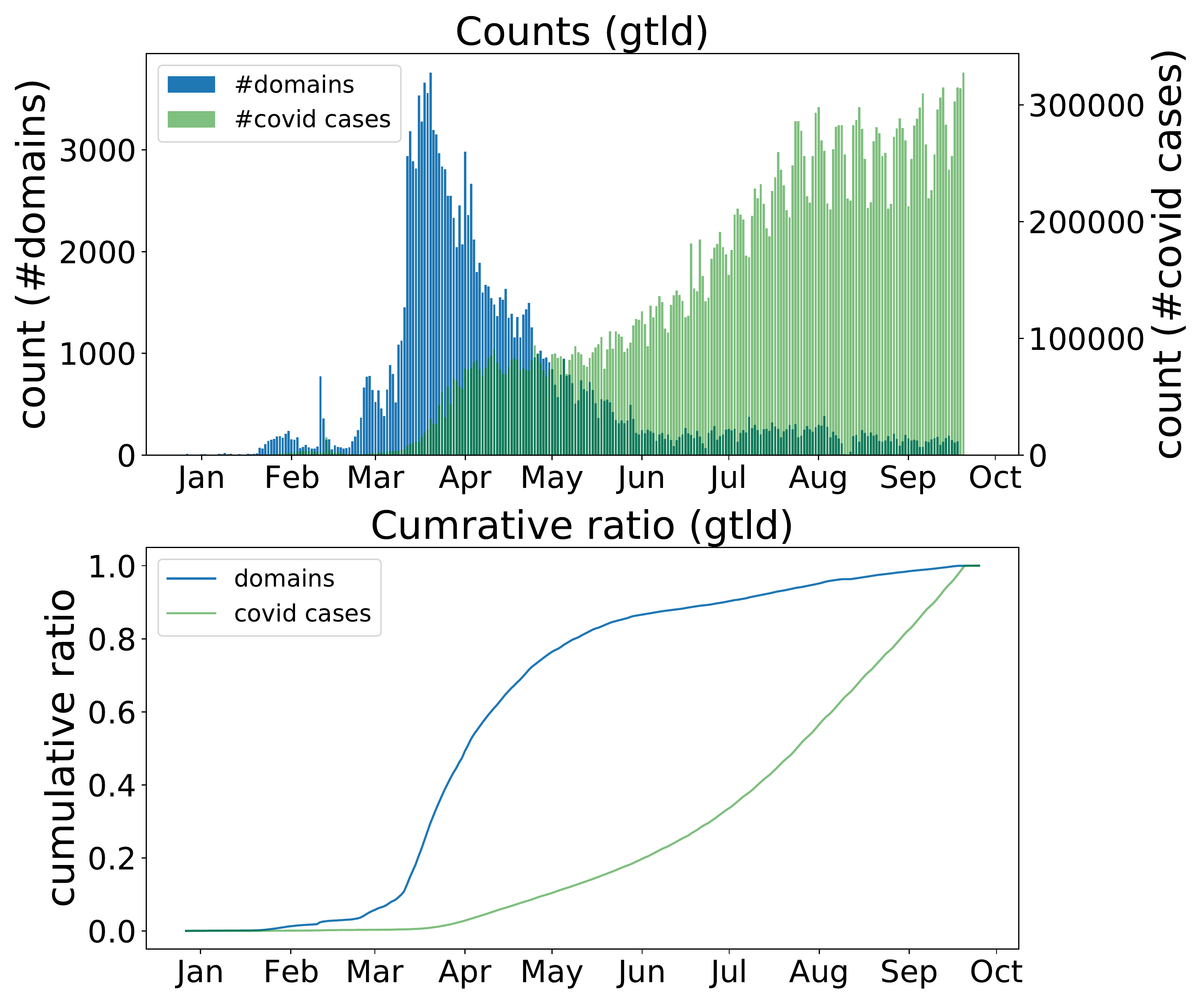}
    \includegraphics[width=0.49\linewidth]{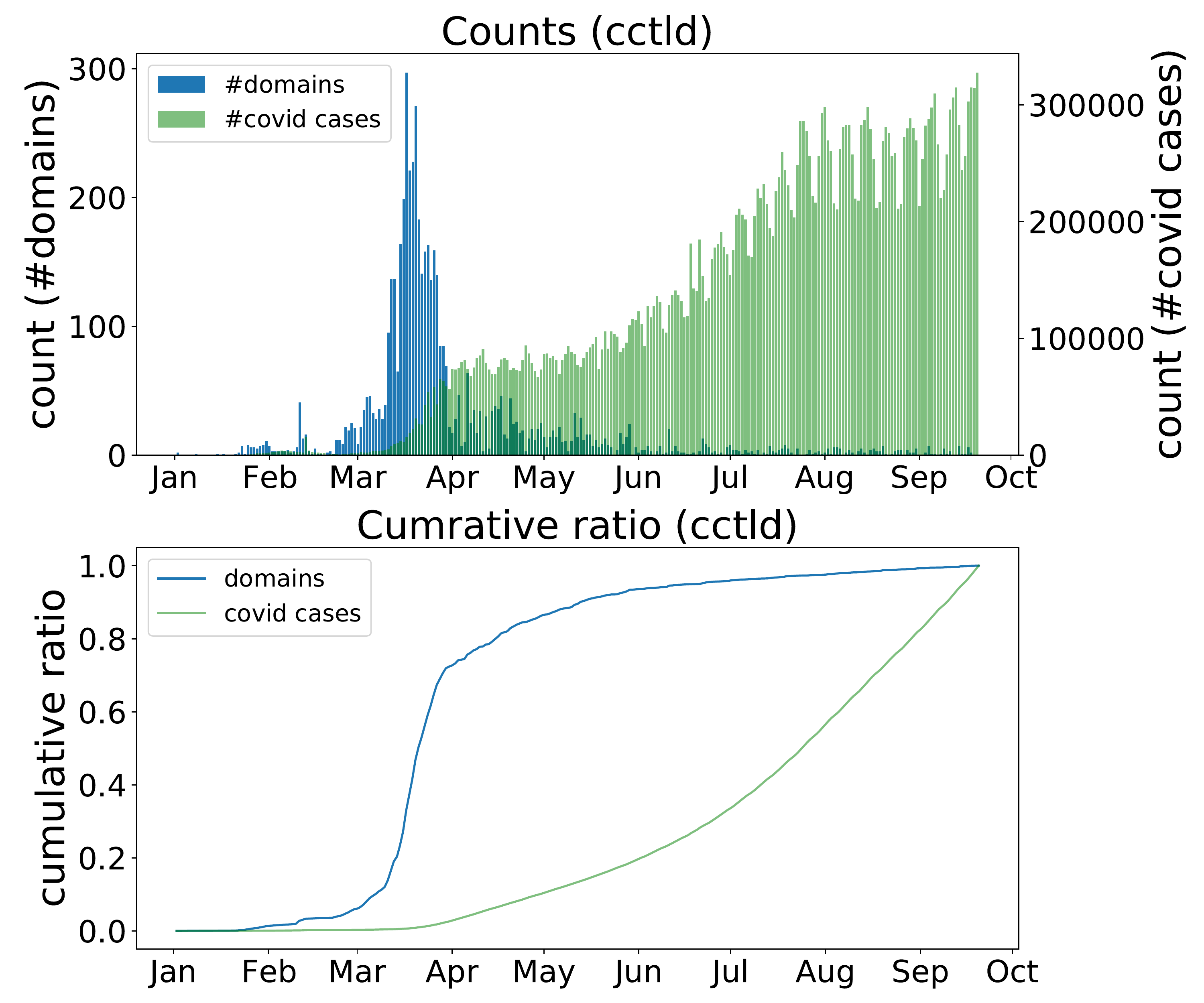}
    \caption{Number of COVID-19 infections and Cov19doms registrations over time. Cases for gTLD (left) and ccTLD (right).}
    \label{fig:time-all}
\end{figure*}
\begin{figure*}[tbp]
\centering
    \includegraphics[width=0.49\linewidth]{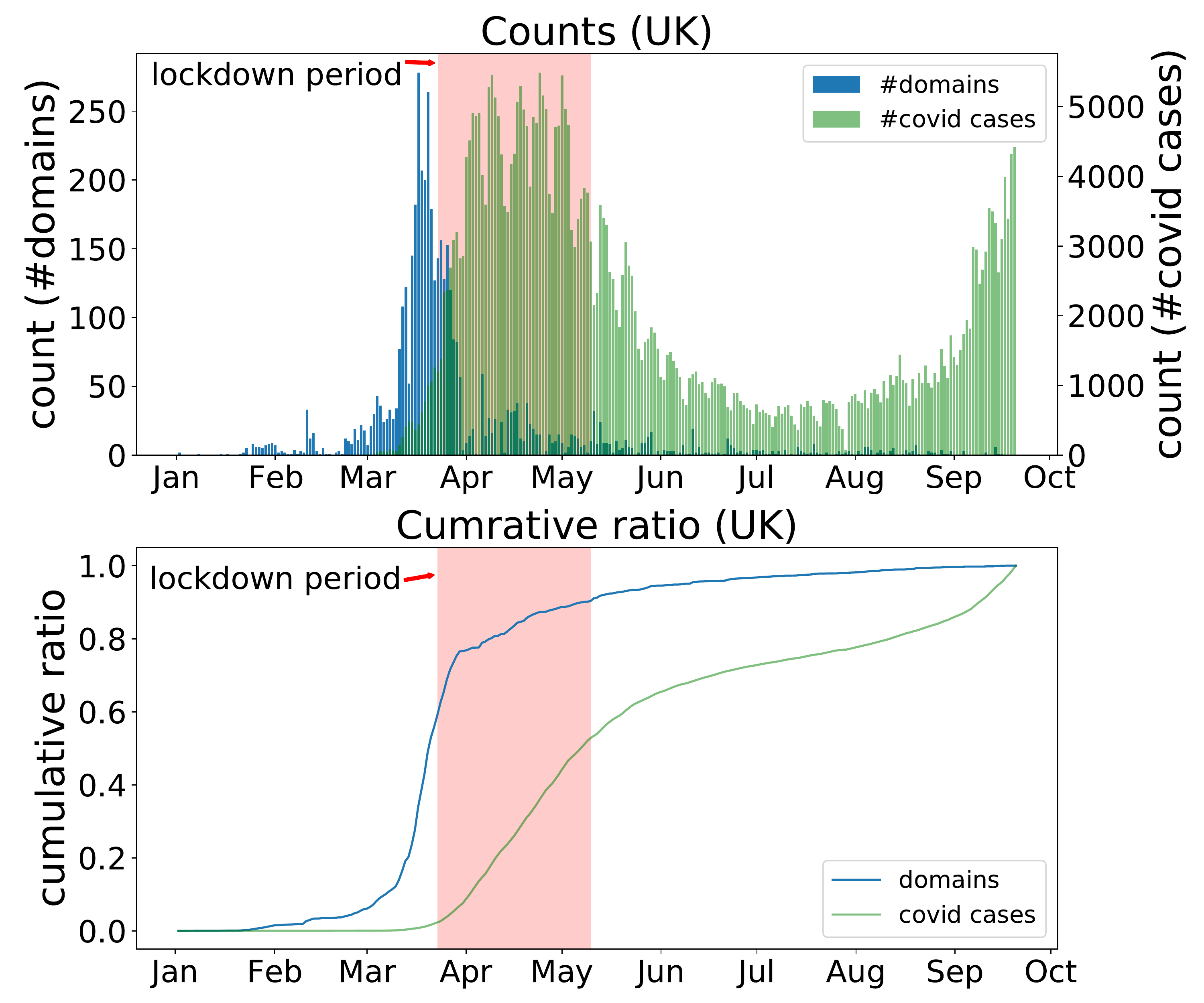}
    \includegraphics[width=0.49\linewidth]{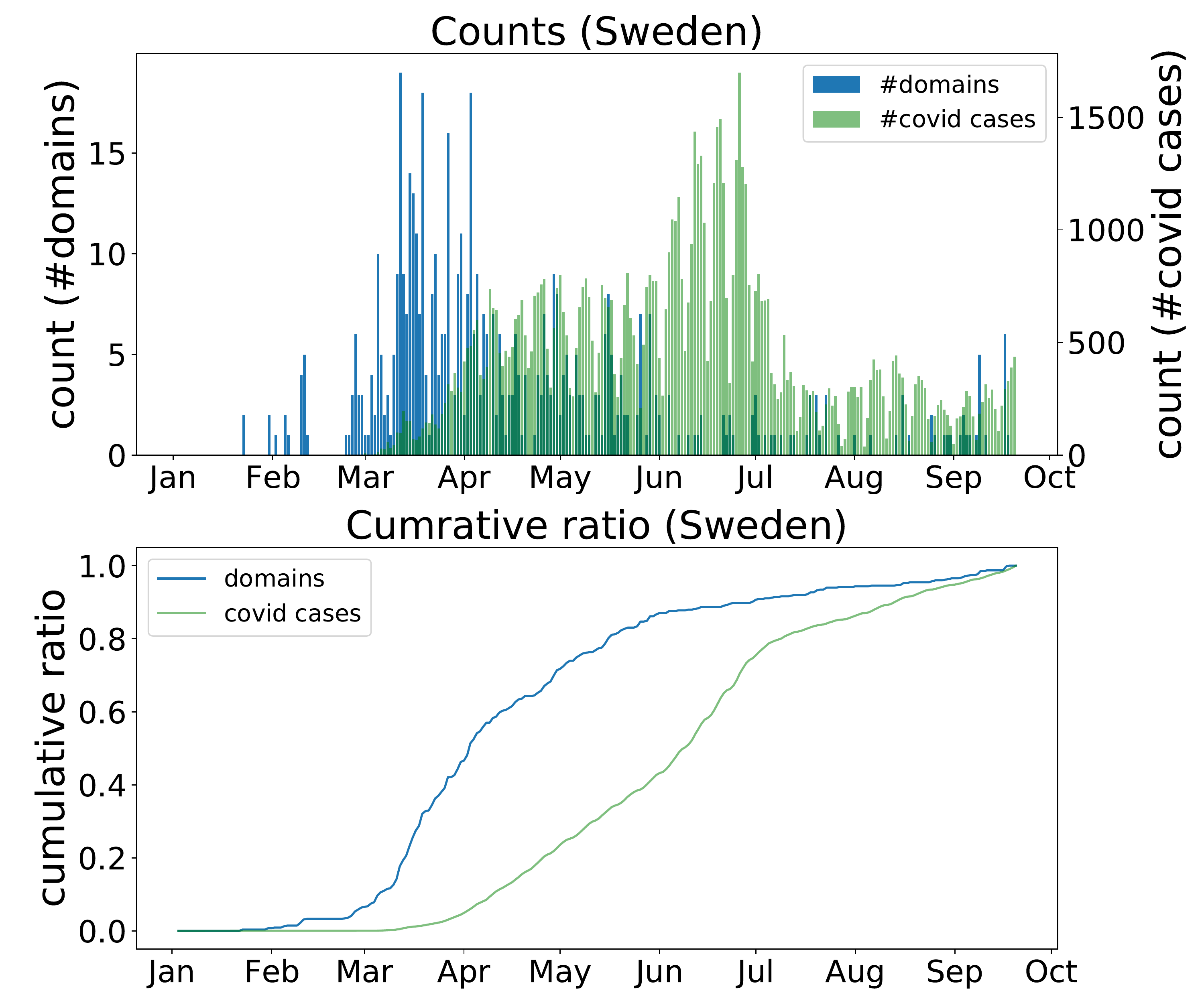}
    \caption{Number of COVID-19 infections and Cov19doms registrations over time. Cases for UK (left) and SE (right). In Sweden, no lockdown enforcement was taken.}
    \label{fig:time-se-uk}
\end{figure*}

First, we investigate the time series of new registrations of Cov19doms and the number of new COVID-19 infections worldwide.
We take all Cov19doms and split them into groups of gTLDs (e.g., \texttt{.com}) and ccTLDs (e.g., \texttt{.uk}).
We obtained information on the number of COVID-19 infections from the official WHO website~\cite{who-covid19}.
Figure~\ref{fig:time-all} shows the time series of the number of new COVID-19 infections and the number of new registrations in Cov19doms (gTLD and ccTLD).
These figures show that similar to the situation of the COVID-19 pandemic outbreak around the world around April 2020, Cov19doms saw a significant increase in its new registrations as well.
Surprisingly, the number of new domain name registrations peaked around March 2020, about a month ahead of the peak in the number of new COVID-19 infections.
Subsequently, the number of new registrations of Cov19doms has reached a stable daily registration rate, but the number of COVID-19 infections has still increasing as of October 2020.

Second, we focus on the Cov19doms of ccTLDs and investigate the relationship between the number of new registrations of Cov19doms per ccTLD and the number of new COVID-19 infections in the country corresponding to the ccTLD over time.
Our Cov19doms data included only four ccTLDs: United Kingdom (\texttt{.uk}), Sweden (\texttt{.se}), Niue (\texttt{.nu}), and Australia (\texttt{.au}).
We excluded \texttt{.nu}, for which no information on the number of WHO infected people existed from there, and \texttt{.au}, for which we were unable to obtain the full domain name registration date from the WHOIS data, and conducted a survey of 4,766 \texttt{.uk} and 549 \texttt{.se} Cov19doms.
Figure~\ref{fig:time-se-uk} shows the time-series change in the number of new infections of Cov19doms and COVID-19 in the UK and Sweden, respectively.
Since the lockdown was implemented in the UK, the period is also shown in .uk graph.
In both cases, Cov19doms registrations tend to be more likely to be ahead of the COVID-19 infection explosion.
Furthermore, we find that registration of Cov19doms moves faster and clearer in the UK than in the Sweden case.

Our results obtained so far above indicate two things: (1) events like COVID-19 that affect so many people's lives will create a massive demand for domain names and (2) people are anticipating such demand and taking the action of registering domain names at an amazingly early stage.
In subsequent sections, we will clarify for what purpose people are registering these Cov19doms.

\subsection{Understanding the Usage of Cov19doms}
\label{sec:categorization}

In general, automatic website classification is not an easy task as the modern web is composed of rich and complex multimedia, making it difficult to automatically analyze its contents using simple data processing scheme.
Therefore, instead of fully automating the website classification process, this work adopted manual inspection to ensure the quality of the classification.
However, the number of Cov19doms we have collected is so large that it is infeasible to inspect them all manually.
Therefore, we took the approach of applying random sampling to reduce the number of domains/websites to be analyzed. 
As shown in Figure~\ref{fig:measurement_overview}, we randomly sampled 10K of websites from 77K active Cov19doms websites to reduce the number of samples to be classified by human.
For the 10K of randomly sampled Cov19doms websites, we took the following two-stage approach.

In the first stage, we aim to systematically classify websites into the following categories: {\em Empty}, {\em Error}, {\em Parked}, {\em Hosted}, and Has content, where 
{\em Empty} represents cases in which HTTP/HTTPS requests were responded to, but the data was empty, 
{\em Error} represents the websites responded with error codes such as 404 or 501, 
{\em Parked} represents the domain parking websites, 
{\em Hosted} represents cases where the domain name has been purchased, but the website only shows the initial page after installation of Apache, WordPress, etc, and
{\em Has content} represents the remaining Cov19doms  websites that have some content.
In the second stage, three evaluators manually classify the websites classified as ``Has content.''
In the following, we present the details of the analysis to be performed at each stage and the results obtained.

\subsubsection{Stage 1: Systematic Classification}

We classify websites into the five classes defined above based on HTTP/HTTPS response codes and screenshot information.
Among the five classes, the classification of empty and error is simple.
They can be classified by analyzing the size of the data retrieved and the response code.
For the remaining classes parking and hosted, we use cluster analysis.
For parking, we could use domain name registrar information in some cases, however, our preliminary study shows that we cannot do a comprehensive study due to the existence of so many different domain parking companies.
The key idea is that the majority of websites that are accessed for parking and hosted are similar in appearance.
Therefore, we apply cluster analysis to the screenshot images and classify the websites by determining whether each cluster is {\em Parked} or {\em Hosted} or {\em Has content}.
With this approach, we can streamline the classification.

To perform clustering of screenshot images, we need to calculate the distance between images; i.e., it is necessary to compute the similarity of images.  
There are several methods for computing the similarity of images, and in this paper, we adopt the perceptual hash (pHash)~\cite{phash}, which computes close hash values for two similar images.
pHash is widely used to discover copyright infringement and is known to be effective in discovering resemblances to certain images.

We first accessed 10K of randomly sampled active websites and extract HTML, screenshots, and other metadata by navigating Google Chrome\footnote{We used the version of 81.0.4044.129.} using Selenium~\cite{selenium}.
The language was set to English, and the User-Agent was set to Windows 10 Google Chrome.
To not halm the websites set to be investigated, access to the IP address corresponding to each Cov19dom is limited twice (HTTP and HTTPS).
Next, we computed the pHash values for the 10,000 screenshots we collected, using imageHash~\cite{imagehash}.
We then grouped the corresponding Cov19doms with the same value of pHash and HTTP status code pairs into the same cluster. 
Table~\ref{tab:systematic-classification} presents the classification result of the Cov19dom websites.
From the table, we can see that many of the Cov19doms websites resulted in either domain parking or errors, and that 40\% of the websites (classified as ``Has content'') requires detailed manual inspection.
We note that 60\% of the websites categorized as other than ``Has content'' do not currently provide any useful content, however, they might start providing some content in the future, so we need to pay attention to them.
In the following, we will classify the websites categorized as ``Has content.''

\begin{table}[tbp]
    \centering
    \normalsize
    \caption{Result of Systematic Classification.}
    \label{tab:systematic-classification}
    {\renewcommand{\arraystretch}{0.9}
    \begin{tabular}{l|rr}
        \hline
        Category & \# of active websites & fraction (\%) \\
        \hline\hline
        Empty     & 609 &  6.1 \\
        Error     & 1,663 &  16.6 \\
        Parked        & 2,138 & 21.4 \\
        Hosted        & 1,402 & 14.0  \\
        \hline
        Has contents   & 4,188 &  41.9  \\
        \hline\hline
        Total    & 10,000 & 100.0   \\
        \hline
    \end{tabular}
    }
\end{table}

\subsubsection{Stage 2: Manual Classification}
In the second stage, we will classify the Cov19doms websites marked as ``Has content'' in the Stage 1. 
Since 4K of websites are too many to analyze manually, further random sampling is performed and 1,000 general websites will be carefully classified by three evaluators.
Through the Stage 1 classification, the classification categories for Stage 2 were predetermined and provided to the evaluators with detailed explanations.
Figure~\ref{fig:classify-tool} presents a screenshot image of a tool developed by the authors to help evaluators efficiently classify websites.
Although the evaluators made a classification based on screenshot image and metadata, there are cases that cannot necessarily be determined by screenshot or metadata.
For example, if the evaluators could not understand the language used in the web content, they also leveraged external resources such as a search engine.

 \begin{figure*}[tbp]
     \centering
     \includegraphics[width=0.98\linewidth]{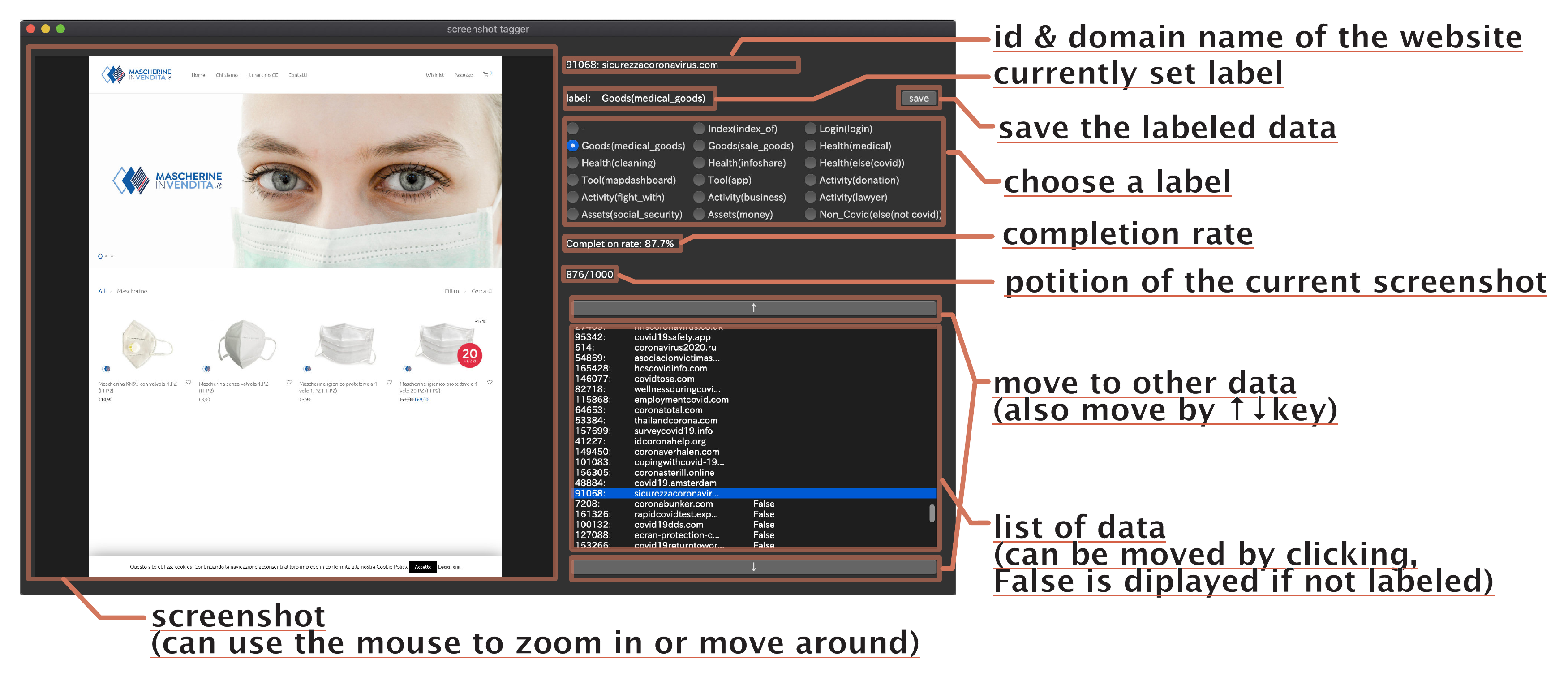}
     \caption{A screenshot of the website classification tool we developed for our analysis.}
     \label{fig:classify-tool}
 \end{figure*}

Three evaluators used the tool to classify 1,000 of websites marked as  ``General'' taking 4.8 hour on average, resulting in 477 websites where the three evaluators agreed, 423 websites where the two evaluators agreed, and 100 websites where they all disagreed.
That is, for 90\% of the websites, at least two evaluators' classification results were consistent.
The result of calculating the Fleiss' kappa coefficient, which is a quantitative measure of inter-rater agreement, was 0.50, which can be interpreted as moderate agreement~\cite{landis1977measurement}.
The results of the interviews with the three evaluators revealed that the primary reason for the disagreement was the difference in the decisions they made when they were unsure of their classification. One of the evaluators reported that he categorized all of his confusion as ``activities for COVID-19.''
Discrepancies in judgments also arose because of the existence of websites that could be classified into multiple categories. For example, a website that displays medical products (masks, face shields) may be categorized as both Health and Sales.
Apart from such discrepancies, the classifications were generally consistent and the manual classification results can be considered reasonable.
In the final classification, a majority vote was adopted.
Websites with discrepancies between the three evaluators' classification results were marked as ``Unknown.''

Table~\ref{tab:manual-classification} presents the classification results.
About 70\% of the websites were related to COVID-19.
Those websites were medical services, selling products, providing COVID-19 information such as apps, maps and dashboards, supporting people's activities related to COVID-19, and social security.
As would be expected from the nature of COVID-19, the majority of the websites (40\%) were medical-related.
Many of these health-related websites are critical sources of information in countering COVID-19 pandemic and should never be blocked.
The remaining 30\% were completely irrelevant websites, websites with no content displayed, and ``Unknown,'' which we defined earlier.　

\begin{table*}
\centering
\normalsize
\begin{threeparttable}[!t]
    \caption{Results of Manual Classification.}
    \label{tab:manual-classification}
    {
    \begin{tabular}{l|l|r}
    \hline
    Category & Description & \#sites \\
    \hline\hline
 Health & Websites providing information on health\tnote{$^\ast$}  & 405\\
 Sales & Websites selling products related to COVID-19 & 109\\
 Tools & Websites providing apps/maps/dashboards of COVID-19 & 123\\
 Activities & Websites dedicated to people's activities to address COVID-19\tnote{$^{\ast\ast}$} & 72\\
 Social Security & Websites regarding to social security  & 2\\
 \hline
  Unrelated & Websites unrelated to COVID-19  & 139\\
  Login & Websites showing a login page  & 26\\
  Index of & Websites showing the ``Index of /'' page  & 24\\
\hline
 Unknown &Websites with discrepancies between the evaluators' classifications & 100\\ 
   \hline\hline
 Total &--  & 1,000\\
 \hline
    \end{tabular}
    }
    \begin{tablenotes}
        \item[$^\ast$] hospitals, infection testing, sterilization, and other health-related topics.
        \item[$^{\ast\ast}$] Fundraising, volunteering, business, and political movements regarding COVID-19.
    \end{tablenotes}
\end{threeparttable}
\end{table*}

\subsection{Malicious Activities Using Cov19doms}
\label{sec:malicious}

Finally, we investigate whether Cov19dom websites were involved in any malicious activities.
To achieve this goal, we utilize VirusTotal, a large-scale online virus scanning service.
As shown in Figure~\ref{fig:measurement_overview}, we target 10K of active websites that had Cov19doms.
Of the 10K websites, 6,362 of the websites were detected as malicious by at least one or more scanners.
This is an alarming number, but when we analyzed the detection results, we found that one online scanner detected 6,256 websites as malicious, and that the majority of them (about 98.7\%) were classified as phishing sites.
Although we cannot determine from our data whether or not these detections were correct, the result does suggest that there may be a non-negligible number of malicious sites that use Cov19doms.
On the other hand, one of the reasons why online scanners may falsely detect Cov19doms as a phishing site is likely to be naïve detection using keyword matching.
For example, a scheme that increases the probability of detecting a website with a domain name containing the strings corona or covid-19 as a malicious site could be employed.
However, such an approach might have the risk of blocking websites that provide important information about COVID-19. 

To reduce the effect of false positives from individual scanners, we examined websites that were detected as malicious by at least two online scanners. We note that this approach is consistent with the best practice used in many papers that make use of multiple engines/vendors of VirusTotal for the labeling task~\cite{IMC_VT}. As a result, we found that the number was 357, which accounted for roughly 4\% of the active Cov19doms websites.
The detection categories of those detected by two or more online scanners are summarized in Table~\ref{tab:scan}.
Note that a website may be detected as a different category (e.g., phishing site and malware site) by several online scanners . In such cases, the category is decided by majority vote, and if the category is not uniquely determined, the category is marked as ``pending''.
It can be seen that once again, phishing sites have the highest number of detections, but the number of other malicious sites is also very close.

\begin{table}[!t]
    \centering
    \normalsize
    \caption{Breakdown of the detection results.}
    \label{tab:scan}
    {\renewcommand{\arraystretch}{0.9}
    \begin{tabular}{l|rr}
    \hline
    Detected category & \# detections & fraction (\%) \\
    \hline\hline
    Phishing site & 117 & 32.8 \\
    Malicious site & 52 & 14.6 \\
    Malware site & 17 & 4.7 \\
    \hline
    Pending & 171 & 47.9 \\
    \hline\hline
    Total & 357 & 100.0 \\
    \hline
    \end{tabular}
    }
\end{table}

\vspace{-3mm}
\section{Discussion}
\label{sec:discussion}

\subsection{Limitations}

This study aims to understand the Cov19doms in the wild.
In order to ensure the accuracy of the results, two heuristics were applied to extract such domain names, as described in Section~\ref{sec:data}.
The first heuristic was to limit the domain search words to ``covid'' and ``corona.'' 
Such limitation will miss several cases where domain names contain other keywords such as ``virus'' or ``mask,'' which could bring false positives as we discussed.
We also limited our search to the e2LD part; the limitation will eliminate the cases where an FQDN contains the substrings in its hostname.
Another heuristic was to constrain the registration date for domain names.

Our analysis also excluded websites that did not include keywords in their domain name but were COVID-19-related in their website content.
Such websites existed on both malicious and benign sites.
Another limitation that we are aware of is that the URL path is not taken into account when creating a URL from an FQDN. 
We only retrieved web content from the top directory on a website 
in the web-crawling process.
Exploring the URL path might reduce the errors shown in the Table~\ref{tab:systematic-classification}, however we may miss web content if a website does not configure the setting of index file.
Addressing these issues is left for future study.

\subsection{Detecting malicious Cov19doms}
As we have shown in this work, simply using a list of Cov19doms as a blocklist may result in false positives, and this introduces the risk of blocking information that is useful for COVID-19 countermeasures.
In order to determine if a detected Cov19dom is malicious, we need to monitor a domain name when it is being abused and examine the content in a timely manner.
The Trademark Clearinghouse (TMCH) is a global database of trademarks and provides this information to registries and registrars during the domain name registration process to thwart unwanted domain name registrations by third-parties. 
This is effectively used by trademark owners to fight against a trademark infringement using fake domain names. 
Unfortunately, this countermeasure is not effective against domain names piggybacking on global crises including COVID-19, due to the fact that there is no right owner of such corresponding keywords. 
Szurdi and Christin proposed the anti-bulk registration policy such as dynamic pricing to make bulk domain registrations expensive~\cite{Szurdi:WEIS18}, which is a potential countermeasure against bulk-registered COVID-19 domain names.

\subsection{Ethical considerations}
Our study analyzed publicly available DNS records and web content corresponding to the domain names without collecting personally-identifiable information. 
In our web-crawling process, we sent the minimum amount of legitimate requests to websites, i.e., two requests (HTTP and HTTPS) per site, and left them and their users unharmed.
\section{Related Work}
\label{sec:related}

In this section, we present several related works and clarify how our work differs.

\noindent\textbf{Internet Measurement driven by COVID-19.}
Favale et al.~\cite{Favale_2020} analyzed the impact of the lockdown enforcement on a campus network in Italy.
Through analyzing Internet traffic statistics, they revealed that while incoming traffic 
was reduced by a factor of 10 during the lockdown, outgoing traffic increased by 2.5 times, driven by more than 600 daily online classes, with around 16,000 students per day.
They concluded that the campus network infrastructure is robust enough to successfully cope with the drastic changes while maintaining the university operations.
Feldmann et al.~\cite{IMC20-covid-isp} conducted similar analysis using traffic data collected at one ISP, three IXPs, and one educational network. 
They reported on changes in Internet traffic in various perspectives and concluded that the Internet infrastructure has been able to deliver the increased Internet traffic without significant impact.

Candela et al.~\cite{c2020impact} conducted a large-scale analysis of Internet latencies, which could be affected by the increased amount of online activities during the lockdown.
By leveraging the measurement data collected with the RIPE Atlas platform~\cite{ripeatlas},
they analyzed Internet latencies focusing on Italy, where people experienced more than a month of lockdown.
They reported that the increase in online activity led to an increase in the variability of Internet latencies, a trend that intensified in the evening due to the increase in the entertainment 
traffic.

\noindent\textbf{Event-driven domain name registration.}
The strategy of early acquisition of domain names associated with ongoing events has been a well-known approach in the domain name business community. In fact, a patent of such a technique was filed by an Internet domain registrar~\cite{patent-2008}. 
Although event-driven domain registration is a widely known best practice in the domain name business community, to the best of our knowledge, there has been little research on the topic in the research community. 
One of the few available studies is that Coull et al.~\cite{10.1007/978-3-642-15257-3_7} 
derived rules to describe topics, such as ongoing events, from popular Google search queries with the aim of characterizing the registration of speculative domain names and empirically evaluated the feasibility of domain acquisition based on such a method. While they attempted to extract current events using Google search, COVID-19 is a unique phenomenon, and researchers have not had an opportunity to study domain names for such a case.

Tombs et al.~\cite{ambiguity_in_covid_tld} tried to determine the level of credibility of a top-level coronavirus-related website that purport to be government websites, and find out the purpose of non-governmental entity or company register a top-level coronavirus-related domain name by analyzing data collected from 303 websites which domains related to COVID-19 between April 5 and April 6, 2020. 
They found that 80\% of websites presented as government websites cannot be verified the authenticity. Additionally, about 30\% of websites collected had unverified information and nearly half were squatting domains or ``under construction."
Government websites providing critical information about coronaviruses should not be subject to ambiguous in their authenticity , and therefore should not share the top-level domain name space with non-governmental entity or company.
Their findings are important in establishing trusted communication channel between government and  their citizens during this crisis.

\noindent\textbf{Malicious domain names and websites.}
Much research has been conducted on ways to observe the registration and early activity of malicious domain names~\cite{10.1145/2068816.2068842,10.1145/3196494.3196548,10.1145/3278532.3278569}. 
Hao et al.~\cite{10.1145/2068816.2068842}
unveiled that DNS infrastructures and early DNS lookup patterns for a newly registered malicious domain name differ significantly from those with a legitimate domain name. 
Korczynski et al.~\cite{10.1145/3196494.3196548}
collected WHOIS information, web content, and DNS records for corresponding malicious domain names provided from 11 distinct abuse feeds and observed a growing number of spam domains in new gTLDs, indicating a shift from legacy gTLDs to new gTLDs.
We conducted our measurement by referring to the ways practiced in these existing studies.
While these studies analyzed fake domain names containing strings related to brand names having specific owners, our study focuses on domain names containing strings related to generic crisis having no specific owners, which makes it be challenging to distinguish between malicious and legitimate domain names.

There are few academic studies so far on detection of malicious domain names related to COVID-19.
Ispahany and Islam developed a machine learning model using lexical features to detect malicious domain names and examined registered domain names in April 2020~\cite{ispahany2020detecting}. The purpose of our study is not to detect malicious Cov19doms, but to investigate the usage of Cov19doms. 
Furthermore, our study utilized a long-term dataset obtained from the end of December 2019 to the end of September 2020.

\section{Conclusion}
\label{sec:summary}

Through the analysis of 167K of Cov19doms we collected, we found that a month before the global COVID-19 pandemic hit in April 2020, there was a flood of domain name registrations.
This phenomenon can be attributed to a variety of people registering domain names for the purpose of COVID-19 countermeasures, speculative domain name business, or to generate phishing sites, as they predicted the high impact of COVID-19.
Such a global, high-impact phenomenon is unprecedented in the past and is a remarkable event from the perspective of Internet measurement.
In conventional measures against the registration of unwanted domain names targeting brands, distinguishing between an original domain name and a fake domain name has been relatively straightforward since the brand owner has been determined.
In the case of the Cov19doms, on the other hand, there is no concept of a brand owner, and many different players have registered Cov19doms to benefit society.
Therefore, it is not feasible to apply traditional domain name analysis methods.
As this study revealed, majority of Cov19doms (about 60\%) are not active. Even if Cov19doms are uesd for active websites, many of them are parked or hosted, and it is not clear how these domain names will change in the future.
Addressing these problems is a challenge for the future.
We plan to release our dataset and tools used for our analyses at \texttt{https://github.com/cov19doms/cov19doms}

\bibliographystyle{sp}
\bibliography{ref}

\end{document}